# Biexciton generation rates in CdSe nanorods are length independent

Roi Baer[(a)] and Eran Rabani[(b)]

*(a) Fritz Haber Center for Molecular Dynamics, Chaim Weizmann Institute of Chemistry, The Hebrew University of Jerusalem, Jerusalem 91904 Israel; (b) School of Chemistry, The Raymond and Beverly Sackler Faculty of Sciences, Tel Aviv University, Tel Aviv 69978 Israel.*

Abstract: We study how *shape* affects multiexciton generation (MEG) rates in a semiconducting nanocrystal by considering CdSe nanorods with varying diameters and aspect ratios. The calculations employ an atomistic semiempirical pseudopotential model combined with an efficacious stochastic approach applied to systems containing up to 20,000 atoms. The effect of nanorod diameter and aspect ratio on multiexciton generation rates is analyzed in terms of the scaling of the density of trion states and the scaling of the Coulomb couplings. Both show distinct scaling from spherical nanocrystals leading to a surprising result where the multiexciton generation rates are roughly independent of the nanorod aspect ratio.

Multiexciton generation (MEG) is a process by which several electron-hole pairs are generated from a single photon. In bulk, due to strict selection rules of energy and momentum conservation and rapid nonradiative relaxation, MEG becomes efficient only when the photon energy exceed several times the band gap energy ($E_g$),[1,2] typically at excitation energies $> 5E_g$. In system where translation symmetry is broken, such as nanocrystals (NCs), it was argued that the onset of MEG should be lower,[3] as some of the strict selection rules are removed and exciton relaxation dynamics is altered by quantization effects.

Indeed, experimental studies indicate that the onset of MEG in spherical NCs is lower than bulk when the photon energy is scaled by the band gap energy,[4-16] approaching values of $2E_g < E_{onset} < 3E_g$. This was rationalized by several different theoretical approaches.[17-31] Perhaps the simplest picture is based on impact excitation,[17-21] where the rate of MEG decreases with increasing NC size approaching bulk values as the size of the NC exceeds the Bohr exciton radius. This was explained in terms of scaling of the band gap, Coulomb couplings and density of states with the NC size.[19-21]

Recent observations of MEG in nanowires and nanorods (NRs) have questioned the role of translational symmetry breaking and its effect on MEG yields.[32-34] While the experiments of Gabor *et al.*[32] on carbon nanotubes photodiodes can be rationalized by the presence of internal fields that accelerate the charge carriers to higher energies, resulting in an onset of $E_{onset} = 2E_g$ for MEG,[35] this argument cannot be used to explain the 60% increase in MEG yields in semiconducting NRs compared to spherical NCs.[34] Sandberg *et al.*[34] attributed this increase mainly to a decrease in the electron–hole pair creation energy as the length of the nanorod increases. However, a decrease in electron–hole pair creation energy would also lead to a decrease in the couplings between excitons and biexcitons, which in principle, reduces MEG efficiencies.

In this letter, we address the role of shape of confined semiconductors on MEG yields. We study a series of NRs with aspect ratios ($\zeta = L/D$) ranging from 1 to 10 ($D$ and $L$ are the diameter and length of the NR, respectively) and for different NR diameters. We find that MEG rates depend strongly on the NR *diameter*, but are roughly independent of the NR *aspect ratio*. This surprising result is explained in terms of the scaling of the Coulomb couplings and the density of states with diameter and length of the NRs, which is different compared to spherical NCs.

To calculate the MEG rates for the NRs within the semiempirical pseudopotential model, we adopt the stochastic approach of Baer and Rabani suitable for extremely large systems.[21] In short, the MEG rate is decomposed into a sum of negative and positive trion formation rates from an electron in orbital $\psi_a(\boldsymbol{r})$ and energy $\varepsilon_a$ or a hole in orbital $\psi_i(\boldsymbol{r})$ and energy $\varepsilon_i$, respectively:

$$\Gamma_a^- = \frac{2\pi}{\hbar}\langle W_a^2\rangle \rho_T^-(\varepsilon_a)$$
$$\Gamma_i^+ = \frac{2\pi}{\hbar}\langle W_i^2\rangle \rho_T^+(\varepsilon_i) \tag{1}$$

where

$$\rho_T^-(\varepsilon) = \sum_{j,b,c}\delta\left(\varepsilon - (\varepsilon_b + \varepsilon_c - \varepsilon_j)\right)$$
$$\rho_T^+(\varepsilon) = \sum_{b,j,k}\delta\left(\varepsilon - (\varepsilon_j + \varepsilon_k - \varepsilon_b)\right) \tag{2}$$

are the densities of negative and positive trion states (DOTS) at energy $\varepsilon$, respectively. We use indices $i,j,k,l$ for occupied (hole) state, $a,b,c,d$ for unoccupied states (electron), and $r,s,t,u$ are general indices. In the above equation, $\langle W_a^2\rangle$ and $\langle W_i^2\rangle$ are the trion-weighted average couplings square, given by:

$$\langle W_a^2\rangle = \frac{\sum_{cbj}W_{a;cbj}^2\delta\left(\varepsilon_a - (\varepsilon_b + \varepsilon_c - \varepsilon_j)\right)}{\sum_{cbj}\delta\left(\varepsilon_a - (\varepsilon_b + \varepsilon_c - \varepsilon_j)\right)}$$
$$\langle W_i^2\rangle = \frac{\sum_{jkb}W_{i;jkb}^2\delta\left(\varepsilon_i - (\varepsilon_j + \varepsilon_k - \varepsilon_b)\right)}{\sum_{jkb}\delta\left(\varepsilon_i - (\varepsilon_j + \varepsilon_k - \varepsilon_b)\right)}. \tag{3}$$

The quantities in Eqs. (2) and (3) can be evaluated using a Monte Carlo (MC) procedure, as described in Ref. 21, with the coupling elements given by:

$$W^2_{i;jkb} = 2|V_{jibk} - V_{kijb}|^2 + |V_{kijb}|^2 + |V_{jibk}|^2$$
$$W^2_{a;cbj} = 2|V_{acjb} - V_{jcab}|^2 + |V_{jcab}|^2 + |V_{acjb}|^2, \quad (4)$$

and the Coulomb matrix elements by (we use a dielectric constant $\epsilon = 6$ in the results shown below):

$$V_{rsut} = \iint d^3r\, d^3r' \frac{\psi_r(\mathbf{r})\psi_s(\mathbf{r})\psi_u(\mathbf{r}')\psi_t(\mathbf{r}')}{\epsilon|\mathbf{r}-\mathbf{r}'|}. \quad (5)$$

The single particle states, $\psi_r(\mathbf{r})$, were obtained from a semiempirical pseudopotential models on a real space grid.[36,37] The Coulomb couplings given by the convolution in Eq. (5) were then calculated using fast Fourier transforms.

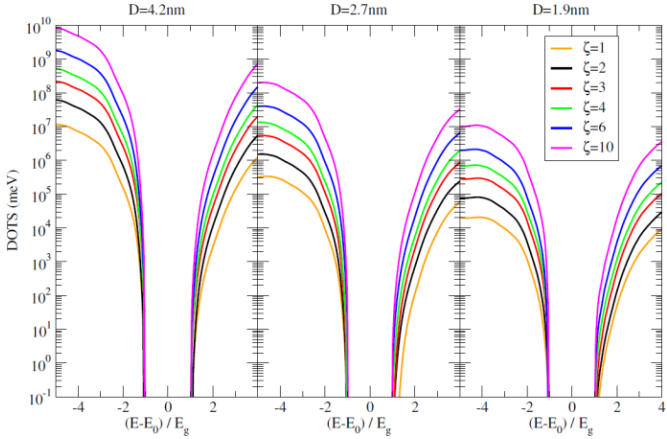

Figure 1: The DOTS versus scaled energy for three different nanorod diameters and for different aspect ratios. Positive (negative) DOTS are plotted for negative (positive) energies measured from the top (bottom) of the valance (conduction) band.

In Figure 1 we plot the negative and positive DOTS defined in Eq. (2) for a series of CdSe NRs at different values of the diameter ($D$) and aspect ratios, $\zeta = L/D$ ($L$ is the length of the NR). The largest system is composed of more than 20,000 Cd and Se atoms. The results were obtained by convoluting the density of states (DOS) calculated by a MC procedure.[21] DOTS for positive trions are plotted for negative energies measured from the top of the valance band while DOTS for the negative trions are plotted for positive energies measured from the bottom of the conduction band. In both cases, the energy is scaled by the gap energy ($E_g$) of the NR. The energy gaps for the largest aspect ratios are $E_g = 2.04, 2.30, 2.59$ eV for $D = 4.2, 2.7, 1.9$ nm, increasing slightly for aspect ratios below $\zeta = 2$.[38]

As expected, the onset of the DOTS appears at energies equal to $\pm E_g$ from the corresponding band edge. The overall shape of the positive or negative DOTS as a function of energy is determined by rather simple scaling considerations. Because the DOTS is a triple convolution of the DOS[21] and the latter is proportional to the volume of the NR, $V = \pi D^3 \zeta$, we expect the DOTS to be proportional to $V^3$. This is indeed the case when comparing the DOTS of different diameters or aspect ratios at a given absolute energy $E$.

In spherical NCs, the DOTS increase as $V^3$ at an absolute energy and as $V^2$ at a scaled energy. This difference in scaling is due to the dependence of the NC energy gap on its diameter. In NRs the difference between scaled and absolute energy is softer. For example, if one holds $D$ constant (for the diameters studied in this work) and changes $\zeta$ then the gap is practically independent of the aspect ratio for $\zeta > 3$. This implies that the DOTS is proportional to $V^3$ even at scaled energy as long as the change in the volume is associated with a change of the length of the NR. However, if $D$ grows while $\zeta$ is fixed then the gap decreases and the DOTS increase as $V^n$ where $2 < n < 3$. In the limit $\zeta \to 1$, we find that $n \to 2$, recovering the scaling of spherical NCs.

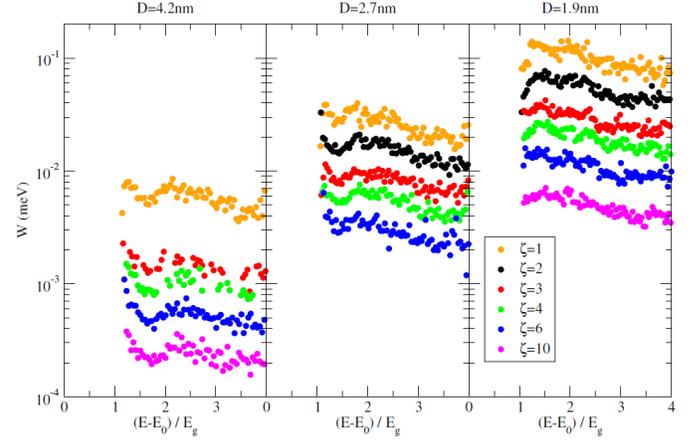

Figure 2: The average value of the Coulomb coupling, $W = \sqrt{\langle W_a^2 \rangle}$, as a function of the electrons scaled energy, measured from the bottom of the conduction band ($E_0$). Different panels correspond to different NR diameter, as indicated. The values of $W$ decrease as the aspect ratio and the diameter of the NR increases.

In Figure 2 we plot the values of $W = \sqrt{\langle W_a^2 \rangle}$ as a function of the scaled energy for different values of the NR aspect ratio and for different NR diameters. In all cases shown, we find that $W$ decreases slightly with increasing energy.[20, 21] Comparing the results for different NR diameters, we find a strong dependence on $D$, where $W$ scales roughly as $D^{-\alpha_D}$ with $\alpha_D \approx 3.5$, very similar to the case of spherical NC.[21] This steep dependence (rather than $D^{-1}$) can be explained in terms of overlapping oscillatory integrals of the electron/hole and corresponding trion states.[39]

The scaling of $W$ with the length of the NR is also given by a power law $L^{-\alpha_L}$, with $\alpha_L \approx 1.3 - 1.5$ depending somewhat on $D$. It lies between the limit of $L^{-3}$ expected when the initial electron/hole wave function ($\psi_{a,i}(\mathbf{r})$) is highly oscillatory[39] and $L^{-1}$ expected when the initial electron/hole wave function is smooth.[39] If the electronic states of a NR are approximated by a particle in a cylinder with transverse angular states (labeled by quantum numbers $n$ and $m$) and longitudinal states (labeled by quantum numbers $k$), the eigenvalues of the electron/hole are then approximated by:

$$E^{e/h}_{nmk} \approx \frac{\hbar^2}{2m^*_{e/h}}\left(\left(\frac{2\phi_{mn}}{D}\right)^2 + \left(\frac{\pi k}{L}\right)^2\right), \quad (6)$$

where $\phi_{mn}$ is the $n$'th root of the $m$'th Bessel function and $m^*_{e/h}$ is the electron/hole effective mass. For each value of $m$ and $n$ there is a manifold of states with different values of $k$, where $k_{max} \propto \zeta$ is the maximal value of $k$ at each manifold. Since $\zeta \leq 10$ in the present calculations, this implies that the initial excitation is characterized by relatively small values of $k$ with wave functions that are quite smooth. This is consistent with the above values of $\alpha_L \approx 1.3 - 1.5$ in between the aforementioned limits.

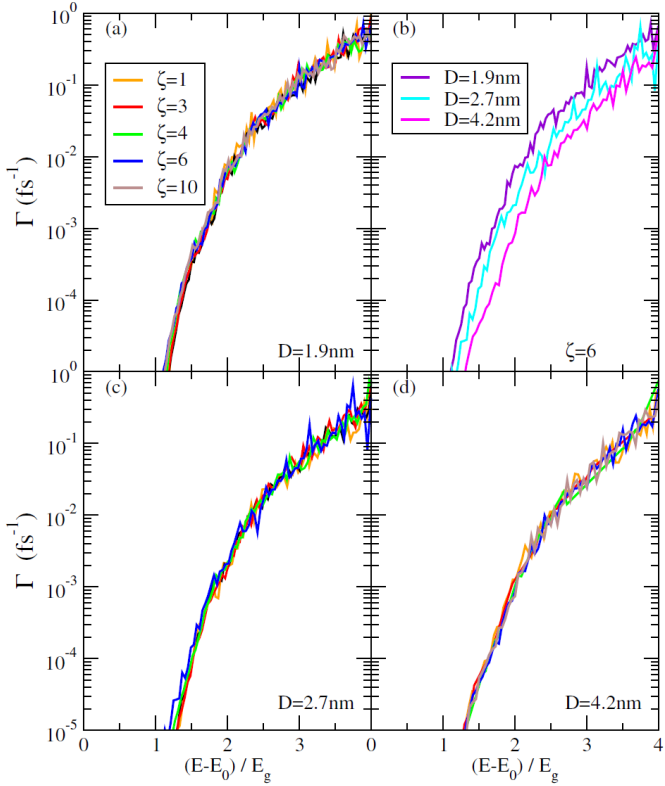

Figure 3: Negative trion formation rates. Panels (a), (c) and (d) show the results for different aspect ratios for NR with diameter $D = 1.9$nm, $D = 2.7$nm, and $D = 4.2$nm, respecitivly. Panel (b) shows the result for NRs with aspect ratio of $\zeta = 6$ for the three diameters.

The scaling dependencies of the trion formation rates are passed from the scaling of the DOTS and the Coulomb coupling $\langle W^2_{i,a} \rangle$ (cf., Eq. (1)). While the DOTS increases as $\zeta^3$ the scaling of $\langle W^2_{i,a} \rangle$ roughly cancels this, leading to trion formation rates that are independent of $\zeta$. In Figure 3 we plot the negative trion formation rates, $\Gamma^-_a$, as a function of the scaled electron energy ($E$) measured from the bottom of the conduction band ($E_0$), for various NR diameters and aspect ratios. The onset of $\Gamma^-_a$ is at the onset energy of $E_0 + E_g$; it increases steeply with energy inheriting this behavior from the DOTS (see Figure 1).

The upper right panel of Figure 3 (panel (b)) shows the negative trion formation rates for different values of the NR diameter $D$. The differences between $\Gamma^-_a$ of NRs with different diameters are clearly resolved, even within the noise level of our MC procedure. We find that at a given scaled energy, $\Gamma^-_a$ decreases roughly as $D^{-1}$. It is interesting to note that the $D^{-1}$ scaling for NRs is fundamentally different from that of NCs,[21] where it scales as $D^{-2}$.

The remaining panels of Figure 3 show $\Gamma^-_a$ for a given NR diameter and for different aspect ratios. *Indeed, we find that the trion formation rates are largely independent of the aspect ratio* $\zeta$, within the noise level of the MC calculations. This surprising result is, however, consistent with the picture developed for the scaling of the DOTS and of the Coulomb couplings.

In summary, we have used an atomistic semiempirical pseudopotential model combined with a stochastic approach to calculate the MEG rates in CdSe NRs of size $\leq 20,000$ atoms and for energies as high as $4E_g$ above/below the corresponding bands. We showed that the MEG rates increase rapidly with energy inheriting this behavior from the DOTS, decrease with NR diameter at a scaled energy, and are roughly independent of the NR aspect ratio within the range $1 \leq \zeta \leq 10$ studied. The latter was rationalized by analyzing the scaling of the Coulomb coupling ($\approx L^{-1.5}, D^{-3.5}$) and the DOTS ($\approx L^3, D^6$). The scaling behavior of the $W$ in NRs is quite different from that of spherical NCs, resulting from the difference in level structure and scaling of the energy gap. While MEG efficiencies depend also on the rate of nonradiative relaxation (which is expected to be independent of $\zeta$ for $\zeta > 2$), our results suggest that these yields are largely independent of the NR aspect ratio. This prediction awaits experimental verification.

**Acknowledgments.** This research was supported by the Israel Science Foundation (grant numbers 611/11, 1020/10) ER would like to thank the Center for Re-Defining Photovoltaic Efficiency Through Molecule Scale Control, an Energy Frontier Research Center funded by the U.S. Department of Energy, Office of Science, Office of Basic Energy Sciences under Award Number DE-SC0001085.